\documentstyle[aps,epsf]{revtex}

\begin{document}


\title{A First-Principles Theory of Bloch Walls in 
Ferromagnets}

\author{J.\ Schwitalla and B.\ L.\ Gy\"orffy}
\address{H. H. Wills Physics Laboratory, University of Bristol, Bristol, U.K.}
\address{Tyndall Ave, Bristol BS8-1TL, U.K.}
\author{L.\ Szunyogh}
\address{Department of Theoretical Physics, Budapest University of 
Technology and Economics, Hungary}
\address{Budafoki u. 8, 1111 Budapest, Hungary}
\address{and Center of Computational Materials Science, Technical University Vienna, Austria}
\address{Gumpendorferstr. 1.a., 1060 Vienna, Austria}
\address{\ }

\date{submitted to PRB, \today}

\maketitle

\begin{abstract}
We present a first principles account of Bloch Walls on the
basis of relativistic Spin-Density Functional Theory in
the Local Density Approximation (LDA). We argue that
calculations based on this description will provide useful
and novel informations about the physics of this important 
element of micro-magnetics. To illustrate the points we 
wish to make we implement the proposed calculations for Fe
and determine the Bloch Wall thickness $\ell_{BW}$.
Moreover we study features of the electronic structure 
which arise due to the presence of the Bloch Wall.
\end{abstract}


\section{Introduction}
Domains of different magnetic orientations spontaneously
nucleate in all magnetic materials in order to minimize 
the magnetostatic energy (dipolar interaction) of the system.
These domains are separated by domain or Bloch Walls where
the magnetization is changing over, rapidly, from one orientation to an
other. In this paper we shall develop a theoretical account 
of these transition regions in terms of mobile electrons and
their spins, that is to say from first principles.

The understanding and control of the domain structure,
magnetic morphology, is central to the effective use of
magnetic materials in many technologies \cite{HS98}.
Moreover, the nature, shape, and motion of Bloch Walls have
always attracted fundamental scientific interest 
\cite{Heis31,Bloch32,LaLi35,Lifs44,Neel44,Kit49,Brown63}.
Nevertheless, all the theoretical discussions so far,
of the entire subject, have been strictly phenomenological.
In the present paper we wish to lay the foundation of a
first-principles approach to the problem.

To start with a tractable problem, henceforth, we shall focus on that 
of a single Bloch Wall. Note that whilst its width, $\ell_{BW}
\sim 10-1000$ $nm$, much larger than the lattice spacing $a$,
it is smallest of all the other length scales such as
the sizes of the domains, $1-1000$ $\mu m$, magnetic
texture, $> 0.1$ $mm$, etc. in the problem.
Thus, by deploying our first-principles methodology, designed
to treat variations in properties on the length-scale $a$,
to Bloch Wall problems,
we are making the natural choice of crossing the smallest 
scale-gap first.

The established theory of Bloch Walls is a topic in 
micro-magnetics \cite{Kit49} and it is wholly
phenomenological. It consists of the equations of
Landau and Lifshitz \cite{LaLi35} as 
generalized by Brown \cite{Brown63}.
For clarity and easy future reference, we summarize here
the part of this theoretical framework relevant to our present
concern. In short, the magnetization density is described
by the vector field $\vec{M}(\vec{r},t)$, which evolves in time
according to the Landau-Lifshitz equation,

\begin{equation}
\frac{\partial \vec{M}}{\partial t}
=
\frac{2\mu_{\rm B}}{\hbar}
\vec{M} \times \frac{\delta {\cal F}[\vec{M}]}{\delta \vec{M}}
\label{eq:ll}
\end{equation}
where $\mu_{\rm B}$ is the Bohr magneton, $\hbar$ is the
Planck constant and ${\cal F}[\vec{M}]$ is a generalized
free-energy functional of $\vec{M}(\vec{r},t)$.
For our present purposes it will be sufficient to take the 
${\cal F}[\vec{M}]$ to be given by the usual Ginzburg-Landau
expansion,

\begin{equation}
{\cal F} = {\cal F}[\vec M]
\approx
\int d\vec r \left\{ {\cal A} \sum_{i=1}^{3} |\nabla M_i|^2 +
\sum_{i,j=1}^{3} {\cal B}_{ij}M_iM_j + ... \right\}
\label{eq:lg}
\end{equation}
where $i$ and $j$ refer to Cartesian axes $x, y$ and $z$,
and the coefficients ${\cal A}$, ${\cal B}_{ij}$ are material
specific parameters which depend on such thermodynamic 
variables as temperature, $T$, and pressure, $p$. Evidently,
the equilibrium magnetization, $\vec{M}_{eq}$, which
minimizes ${\cal F}[\vec{M}]$ is a stationary solution
of Eq.~(\ref{eq:ll}).
The simplest Bloch Wall like solution of these equations
follow from parameterizing the Cartesian components $M_i$
in terms of spherical coordinates $\theta$ and $\phi$ (see Fig.~1.)

\begin{equation}
\label{eq:mspher}
M_{\rm x} = M_0\cos\phi \sin\theta \, , \;
M_{\rm y} = M_0\sin\phi \sin\theta \, , \;
M_{\rm z} = M_0 \cos\theta
\end{equation}
where $M_0$ is a constant equal to the saturation magnetization,
and stopping the expansion in Eq.~(\ref{eq:lg}) at the 
$4^{th}$ order. For a cubic system (${\cal B}_{ij} =
{\cal B} \: \delta_{ij}$) this means

\begin{equation}
\label{eq:fcubic}
{\cal F}(\theta,\phi) = 
\int \left( \alpha  (\nabla \theta(\vec r))^2  +
\sin^2(\theta)(\nabla \phi(\vec r))^2 +
\beta \sin^2\phi(\vec r)\cos^2\phi(\vec r)\right) d\vec r\, ,
\end{equation}
where the new coefficients are defined as 
$\alpha={\cal A} \, M_0^2$ and $\beta={\cal B} \, M_0^4$, respectively.
Taking $\theta=\frac{\pi}{2}$ and $\phi$ a function of the
$z$ coordinate only, it can be readily shown \cite{Kos86}
that, subject to the boundary conditions $\phi(-\infty)=0$
and $\phi(\infty)=\frac{\pi}{2}$, the function

\begin{equation}
\label{eq:soliton}
\phi(z)={\rm arctan}\, \rm{e}^{\sqrt{\frac{\beta}{\alpha}}\,z}
\end{equation}
minimizes ${\cal F}(\theta,\phi)$ in Eq.~(\ref{eq:fcubic}).
In what follows we shall study this simple example of a 90$^{\rm o}$
Bloch Wall from the point of view of mobile electron whose
spin gives rise to such magnetic configuration.

Given the success and the general satisfaction with the
established phenomenological theory outlined above one 
may, at this stage, protest that such microscopic considerations
are not needed. To answer this objection we shall now pause
briefly to argue the case to the contrary. In the interest
of economy we present our arguments as a list of brief
statements:
\begin{itemize}
\item[{\em a.)}]
Firstly, the phenomenological theory assumes that 
$\vec{M}(z)$ rotates as $z$ goes from $-\infty$ to $\infty$
without a change in magnitude. A first-principles theory
should tell us if and when this is true.
\item[{\em b.)}]
Because, as yet, experiments can not tell us otherwise
\cite{Lil50}, the phenomenological theory accepts the form
of $\phi(z)$ in Eq.~(\ref{eq:soliton}) as immutable.
By contrast, the microscopic theory could discover 
situations (materials) where this is not so. The same thing
can be said about deviations of $\theta$ from $\frac{\pi}{2}$
as a function of $z$ or, indeed, the $xy$ dependence of 
both $\theta$ and $\phi$. 
\item[{\em c.)}]
Of course, first-principles calculations not only yield 
$\vec{M}(\vec{r})$ but also the changes in the electronic
structure due to the deviation of $\vec{M}(\vec{r})$ from
its saturation value, namely the presence of a Bloch Wall.
In other words, they describe the electronic structure which 
supports, consistent with, a Bloch Wall. This novel information
becomes available for the first time in the calculations
we shall report.
\item[{\em d.)}]
The change in the electronic states due to the presence of
a Bloch Wall is, or can be viewed as a description of the
electron--Bloch Wall interaction. Clearly, a study of
this interaction will make an important contribution to
understanding the widely observed magneto-transport
phenomena associated with electrons scattering off
Bloch Walls \cite{TNN+99,TZM+99,vHSK+98,vHSB99}.
\item[{\em e.)}]
Inevitably, a magnetic inhomogeneity will distort the
underlying lattice and such distortions give rise to
Bloch Wall--Bloch Wall interactions and pinning of Bloch
Walls by lattice defects like impurities vacancies or grain
boundaries \cite{HS98}. Although we do not deal with
the appropriate generalization of our first-principles
approach, we wish to stress that
it can be readily adopted to address these issues and this
possibility is an important part of the motivation for
perusing it.
\end{itemize}

Finally, we comment on our choice of BCC iron as the host to
the Bloch Wall in our calculations. As is well known
the saturation magnetic moment per atom $\mu_{\rm s}$ at $T=0$
and the effective moment $\mu_{\rm eff}$ deduced from the
Curie constants, measured at $T>T_{\rm C}$, for metallic
ferromagnets are usually not the same. In fact, the deviation
of $\mu_{\rm s}$/$\mu_{\rm eff}$ from 1 can be taken as a good 
measure of how independent the local magnetic moments are
from their relative orientation. Since BCC iron is a 
famously good moment system, namely $\mu_{\rm s} \simeq \mu_{\rm eff}$,
we expect the first-principles calculations to map optimally onto
the phenomenological theory summarized by Eqs.~(\ref{eq:fcubic}) and
(\ref{eq:soliton}). This is indeed what we find and we
take this fact as a validation of our conceptual framework as
well as our numerical procedures for implementing it.
The prize to pay for having done this 'easy' case is
that the importance of all the novel features listed above
({\em a.-e.}) are minimized. Thus, if we disregard these interesting
but small effects, our calculations reduce to a new,
if complicated, way of computing the material dependent
parameters, e.g. the spin-wave stiffness constant $\alpha$
and the anisotropy parameter $\beta$. As these are also
available from other type of first-principles calculations
\cite{LKAG87,KSK91,TJEW95,RSP97}, the comparison of our
results with those of others in the field is a useful
exercise in its own right.

In the next section we shall describe the first-principles
theoretical framework, based on a relativistic, spin-polarized
density functional theory, for our calculations.This is
followed by a section describing the computational procedures.
We shall present and discuss our results in Sec. 4.,
whereas in Sec. 5. we shall evaluate the progress we
have been able to make and the prospects of the
microscopic approach we have advocated.

\section{A first-principles theory of Bloch Walls}
A fully relativistic Density Functional Theory (DFT) \cite{DG90,Strange98},
which includes the dipolar interaction between electrons,
would, presumably, yield an inhomogeneous ground state with
the domain structure determined by the size and shape of
the sample. Clearly, the corresponding calculations are out
of the question and we shall follow the logic of the
phenomenological theory. Namely, we neglect the dipolar
interaction and study a single Bloch Wall engendered by
the constraint that the magnetization $\vec{M}(\vec{r})$
is oriented along two different easy axes at $z \rightarrow \pm \infty$.
This problem is readily encompassed by the spin-polarized 
relativistic Density Functional Theory in the Local Density
Approximation (LDA) as usually applied in solid-state physics 
\cite{DG90,Strange98}.
There are only two features of the way we shall proceed which
deserve further general comment. Firstly, we note that we shall
not be looking for a ground state, as usual, but a lowest
energy state consistent with a constraint which prescribes a
symmetry different from the ground state. Fortunately, as it is
well known, DFT covers this eventuality \cite{DG90,Strange98}.
Secondly, we shall not attempt to address the vexing conceptual
difficulties that arise in connection with imposing constraints
while solving self-consistently the Kohn-Sham-Dirac equation
of the theory \cite{SUW+98}, but in this preliminary exploration of 
the subject implement our strategy in the simplest possible way
as outlined below.

There is no reason to doubt that the established phenomenological
theory \cite{Kos86} in the introduction gets the essential physics of the
Bloch Wall formation right. From the point of view of formulating
a first-principles version of this theory it may be summarized
as follows: the changing orientation of $\vec{M}(z)$, namely the
variation of $\phi(z)$ in Eq.~(\ref{eq:soliton}), across the
Bloch Wall implies an exchange energy cost which is lower the
slower the variation, whereas the anisotropy energy favors a rapid
change from one easy direction, $\phi=0$, to the other,
$\phi=\frac{\pi}{2}$, and the width of the transition region 
of $\phi(z)$ in Eq.~(\ref{eq:soliton}), 
$\ell_{\rm BW}=\pi \sqrt{\frac{\alpha}{\beta}}$, is determined
by the balance of these two tendencies. Indeed, we can bypass
the solution of the Euler-Lagrange equation, which in the case
of minimizing the free-energy ${\cal F}(\theta,\phi)$ in
Eq.~(\ref{eq:fcubic}) is the famous Sine-Gordon equation
whose solution is $\phi(z)$ in Eq.~(\ref{eq:soliton}), 
and take $\phi(z)$ to be a simple function, $\phi_0(\frac{z}{L})$,
which goes from 0 to $\frac{\pi}{2}$ in a distance of $L$
and minimize the corresponding free-energy ${\cal F}(L)$ with
respect to the width $L$. As can be readily shown, by substituting
$\phi_0(\frac{z}{L})$ into Eq.~(\ref{eq:fcubic}), 
for $\theta=\frac{\pi}{2}$

\begin{equation} \label{eq:fsimple}
{\cal F}(L) = \alpha I_1 \frac{1}{L} + \beta I_2 L \;,
\end{equation}
where
\begin{equation}
I_1= \int  \phi_0'(\xi)^2 d\xi \quad {\rm and} \quad 
I_2= \int \sin^2\phi_0(\xi)\cos^2\phi_0(\xi) d\xi \; .
\end{equation}
Clearly, the first term in Eq.~(\ref{eq:fsimple}), proportional
to 1/$L$, is due to the exchange interaction measured by $\alpha$
and the second term, which is proportional to $L$ and the
constant $\beta$, represents the contribution of the magneto-crystalline
anisotropy energy for cubic systems. Furthermore, ${\cal F}(L)$
reaches its minimum at

\begin{equation} \label{eq:lbw}
L_{\rm min} \equiv \ell_{\rm BW} = 
\sqrt{\frac{\alpha \, I_1}{\beta \, I_2}} \; .
\end{equation}
Given that $I_1$ and $I_2$ are simple dimensionless numbers which
parameterize the profile $\phi_0(\frac{z}{L})$,  this result agrees
well with the exact soliton solution for $I_1/I_2 \simeq \pi^2$.

Evidently, the above discussion suggests a rather straightforward
strategy for first-principles calculations:
\begin{itemize}
\item[{\em a.)}]
Assume a profile $\phi_0(\frac{z}{L})$.
\item[{\em b.)}]
Carry out a density functional calculation for the
magnetization $\vec{M}(z)$ constrained to follow the profile
$\phi_0(\frac{z}{L})$ and calculate the total energy or 
grand potential $\Omega(L)$.
\item[{\em c.)}]
Minimize $\Omega(L)$ with respect to $L$.
\end{itemize}
For the profile we shall
always take the particularly simple choice

\begin{equation} \label{eq:LINBW}
\phi_0(\xi) = \left\{
   \begin{array}{ll}
      0      & \xi<0 \\
      \xi\pi/2 & 0\le \xi \le 1 \\
      \pi/2    & \xi \ge 1
   \end{array} \right.
\end{equation}
which is compared to the soliton solution in Fig.~2. 
For this function $I_1=\frac{\pi^2}{4}$ and $I_2=\frac{1}{8}$
and hence if $\Omega(L)$ is given by Eq.~(\ref{eq:fsimple})

\begin{equation} \label{eq:LINomega}
\Omega(L) = \frac{\pi^2}{4} \alpha \frac{1}{L} + 
\frac{1}{8} \beta L \;,
\end{equation}
which implies $\ell_{\rm BW}=\sqrt{2} \pi \sqrt{\frac{\alpha}{\beta}}$.
Note that this Bloch Wall thickness is $\sqrt{2}$ times larger 
than that deduced from Eq.~(\ref{eq:soliton}). Thus the 
Bloch Wall thickness is to some extent an ill-defined concept
in the present 'prescribed profile' approach to the problem.
As will be clear presently, our calculated $\Omega(L)$ can
be fitted by the above functional form very accurately and
hence it determines the coefficients $\alpha$ and $\beta$
from first principles.

For orientation we note that via the Landau-Lifshitz equation,
Eq.~(\ref{eq:ll}), the free-energy functional given in Eq.~(\ref{eq:fcubic})
implies a spin-wave dispersion relation. In the long wave-length limit
this yields

\begin{equation} \label{eq:SWdisp}
\omega_q = \frac{4 \beta \mu_{\rm B}}{M_0} +
\frac{4 \alpha \mu_{\rm B}}{M_0} q^2 \; ,
\end{equation}
thus, the spin-wave stiffness constant

\begin{equation} \label{eq:SWstiffness}
D = \frac{4 \alpha \mu_{\rm B}}{M_0} \; ,
\end{equation}
and the usual cubic (fourth-order) anisotropy constant
\begin{equation} \label{eq:K4}
K_4 = \frac{a^3}{2} \beta \;.
\end{equation}
As expected, in the presence of magneto-crystalline anisotropy the 
spin-wave spectrum is gaped.

Without any account of dipolar interactions, in non-relativistic quantum
mechanics the orientation of the magnetization is independent 
from the crystal axis. As is clear from the above discussion, 
under this circumstance ($\beta = 0$) there is no Bloch Wall
of finite width. Thus, our calculation of $\Omega(L)$ must include 
spin-orbit coupling which is the major source of
magneto-crystalline anisotropy in transition  metals \cite{Bruno93}.
Rather than treating the spin-orbit interaction in perturbation
theory, we use a fully relativistic spin-polarized Density 
Functional Theory \cite{DG90,Strange98}.

The natural variable in the microscopic theory is the 
magnetization averaged over a unit cell
\begin{equation} \label{eq:mi}
\vec{m}_i = \int_{V_i} d^3r \vec{m}(\vec{r}) \; ,
\end{equation}
where $V_i$ is the volume of the $i$th unit cell. We may then define
the local orientation $\vec{e}_i$ by
\begin{equation} \label{eq:ei}
\vec{e}_i = \vec{m}_i / m_i \; ,
\end{equation}
where $m_i$ denotes the length of the vector $\vec{m}_i$.
It is this effective local orientation that we wish
to constrain to follow the discrete version of the profile
in Eq.~(\ref{eq:LINBW}). Namely, we assume that $\vec{e}_p$
is the same within an atomic plane with position $z_p$ and take
\begin{equation} \label{eq:elayer}
\vec{e}_p = \cos \phi_p \: \widehat{x}
          + \sin \phi_p \: \widehat{y} \; ,
\end{equation}
where for a 90$^{\rm o}$ Bloch Wall, which is $N$ layer thick,
\begin{equation} \label{eq:phip}
\phi_p = p \frac{\pi}{2N} \quad (p=1,\dots,N) \; . 
\end{equation}
Thus, we have to solve the Kohn-Sham-Dirac equation of DFT
for the circumstance where the orientation of the magnetization, $\phi_p$,
is changing from layer to layer within a slab of $N$ atomic layers
and is uniformally $0$ and $\pi$/2 to the left and right,
respectively, of the slab as shown in Fig.~2.
Evidently, the relativistic spin-polarized Screened-KKR method
\cite{SUW95}, which treats just such geometries, is ideal for
tackling this problem. Fortunately, the code which implements
this method scales linearly with $N$ and,
as we shall demonstrate presently, can handle 400-500 layers
with readily available computer power.

\section{The computational method}
We performed calculations proposed in the previous section
by using the spin-polarized
relativistic screened Korringa-Kohn-Rostoker (SPR-SKKR) method 
\cite{SUW95}. Although the method is by now well-established
(see also Refs. \cite{SUW+94,ZDU+95}), for completeness 
and since we also made developments specific 
to the Bloch Wall problem, we briefly outline our computational strategy. 
For an ensemble of individual scatterers the multiple scattering 
theory (MST) yields the one-electron Green function for an arbitrary
complex energy $z$ and real space coordinates $\vec{r}$, $\vec{r}'$ 

\begin{equation} \label{eq:greenf}
{\cal G}(z;\vec{r} ,\vec{r} ') = \sum_{QQ'}Z^{n}_{Q}(z,\vec{r}_{n})\:
\tau^{nm}_{QQ'}(z) \:
Z^{m}_{Q'}(z,\vec{r}_{m} ')^{\dagger}-
\end{equation}
$$ - \delta_{nm} \sum_{Q}\left\{
J^{n}_{Q}(z,\vec{r}_{n}) Z^{n}_{Q}(z,\vec{r}_{n}')^{\dagger}
\Theta(r_{n}-r_{n}')+
Z^{n}_{Q}(z,\vec{r}_{n}) J^{n}_{Q}(z,\vec{r}_{n}')^{\dagger}
\Theta(r_{n}'-r_{n})\right\} \ , $$
where $n$ and $m$ label two specific sites of position vectors 
$\vec{R}_n$ and $\vec{R}'_n$, while
$Q$ and $Q'$ denote pairs of angular momentum
quantum-numbers $(\kappa,\mu)$ and $(\kappa',\mu')$, respectively.
The functions $Z^{n}_{Q}(z,\vec{r}_{n})$ and $J^{n}_{Q}(z,\vec{r}_{n})$,
regular and irregular at $\vec{r}_{n} \equiv \vec{r}-\vec{R}_n=0$,
respectively are properly normalized solutions of the Kohn-Sham-Dirac
equation related to a single, finite-ranged potential well for which
we now assume a spherically symmetric effective potential,
$V^{\rm eff}_n(r)$, and effective field,  $B^{\rm eff}_n(r)$, 
pointing along a local (positive) $z$ coordinate axis. In short,

\begin{equation} \label{eq:ssKSD}
\left( c \vec{\alpha} \vec{p} + (\beta - I_4) mc^2 + 
I_4 V^{\rm eff}_n(r) + \beta \Sigma_z B^{\rm eff}_n(r)
- z I_4 \right) 
\left\{ \begin{array}{c} Z^{n}_{Q}(z,\vec{r}) \\
                        J^{n}_{Q}(z,\vec{r})
        \end{array} \right\} = 0 \; ,
\end{equation}
where 
$$\vec{\alpha} = 
\left( \begin{array}{ccc} 0 && \vec{\sigma} \\ \vec{\sigma} && 0 
       \end{array} \right) 
\quad
\beta = \left( \begin{array}{cc} I_2 & 0 \\ 0 & -I_2 \end{array} \right)
\quad
\Sigma_z = \left( \begin{array}{cc} \sigma_z & 0 \\ 0 & \sigma_z
                  \end{array} \right)
\quad
I_4 = \left( \begin{array}{cc} I_2 & 0 \\ 0 & I_2 \end{array} \right)
$$

\medskip
\noindent
with the usual Pauli matrices $\vec{\sigma}$ and the 
two-dimensional unity matrix $I_2$.
The numerical solution of the above equation with a
corresponding expression for the so-called single-site {\em t}-matrix
was originally given in Refs. \cite{FR83,SSG84}. A particular feature 
resulting from the approach they used is that, although
the {\em t}-matrix has necessarily off-diagonal elements, no coupling
between different $\ell$ values is present. 

We further simplify the problem by adopting the
atomic sphere approximation (ASA) in which the
volume of the sphere is taken equal to the volume of the 
corresponding Wigner-Seitz cell. Clearly, ASA deals with
overlapping potentials which is, in strict sense, prohibited within
the MST, however, mostly for inhomogeneous
systems, this approach is widely used, since on one hand it gives a better 
description of the interstitial region than the muffin-tin approach,
on the other hand it is conceptually much simpler than a full potential
description.
As implied above, our approach allows the orientation of the magnetization 
to vary from site to site as required in the case of a Bloch Wall
where we keep the orientation to be constant within the atomic planes 
but let it rotate from layer to layer around 
the global $z$ direction. Therefore, 
in each layer there will be a rotation $R$, corresponding to 
$\theta=\frac{\pi}{2}$ and $\phi=\phi_p$ (see Eq.~(\ref{eq:phip})), 
which relates the $t$-matrix in the local to that in 
the global coordinate system as follows

\begin{equation} \label{eq:trot}
\underline{t}_{glob}(z)=
\underline{D}(R) 
\underline{t}_{loc}(z) 
\underline{D}(R)^\dagger \; ,
\end{equation}
where $\underline{D}(R)$ denotes a matrix containing, block-wise, 
the irreducible representations of $R$.

As the solution of the above single-site problem 
is relatively easy and universal in applications of MST,
the evaluation of the scattering path operator (SPO),
$\tau_{QQ'}^{nm}(z)$ in Eq.~(\ref{eq:greenf}), is, essentially,
the main difficulty in such calculations.
The geometrical arrangement of the scatterers involved in the system is
put, in MST, into the so-called structure constants,
$G^{nm}_{QQ'}(z)=G_{QQ'}(z;\vec{R}_n,\vec{R}_m)$
($\underline{G}^{nm}(z)=\{ G^{nm}_{QQ'}(z) \}$), which for the
relativistic case can be obtained by a transformation in terms
of the Clebsh-Gordon coefficients from its non-relativistic
counterpart (see e.g. Ref. \cite{Wein90}).
By defining the corresponding matrices having both site and
angular momentum indices \cite{Wein90}
$${\bf t}(z) = \{ \underline{t}^n(z) \: \delta_{nm} \} \,, \;
  {\bf G}(z) = \{ \underline{G}^{nm}(z) \} \,, \; 
  \mbox{\boldmath $\tau$}(z) = \{ \underline{\tau}^{nm}(z) \} \;,  $$
the SPO is given by the following matrix inversion

\begin{equation} \label{eq:rsSPO}
\mbox{\boldmath $\tau$}(z) =
\left( {\bf t}(z)^{-1} - {\bf G}(z) \right)^{-1} \;.
\end{equation}

A particular problem arises from the fact that $\underline{G}^{nm}(z)$
is long-ranged, therefore, the inversion in Eq.~(\ref{eq:rsSPO})
can not be directly performed. For a system with three-dimensional
periodicity the problem can be exactly handled by making use of the
lattice Fourier-transformation, which splits Eq.~(\ref{eq:rsSPO})
into the corresponding $\vec{k}$-projections which has to be
solved in angular momentum space only. The Bloch Wall problem
we deal with exhibits, however, two-dimensional (2D) periodicity in the
$(x,y)$ plane, while in the $z$ direction the translational symmetry
is broken due to the variation of the orientation of the magnetization.
Thus employing two-dimensional lattice Fourier-transforms

\begin{equation} \label{eq:2DFT}
\underline{G}^{pq}(z; \vec{k}_\parallel) = \sum_{\vec{R}_\parallel}
e^{\imath \vec{k}_\parallel \vec{R}_\parallel}
\underline{G}(z;\vec{C}_p + \vec{R}_\parallel, \vec{C}_q) \; ,
\end{equation}
where $p$ and $q$ denote atomic layers, generated by $\vec{C}_p$
and $\vec{C}_q$, respectively, $\vec{R}_\parallel$ are 2D lattice vectors
and $\vec{k}_\parallel$ is a vector in the first 2D Brillouin-zone (BZ),
and the new matrix notation in terms of layer indices
$${\bf t}(z) = \{ \underline{t}^{p}(z) \: \delta_{pq} \} \, , \;
{\bf G}(z; \vec{k}_\parallel) =
\{ \underline{G}^{pq}(z; \vec{k}_\parallel) \} \, , \;
\mbox{\boldmath $\tau$}(z; \vec{k}_\parallel) =
\{ \underline{\tau}^{pq}(z; \vec{k}_\parallel) \} \, , $$
one can write
\begin{equation} \label{eq:2DSPO}
\mbox{\boldmath $\tau$}(z; \vec{k}_\parallel) =
\left( {\bf t}(z)^{-1} - {\bf G}(z; \vec{k}_\parallel) \right)^{-1} \; .
\end{equation}

Evaluating the matrix inversion in 
Eq.~(\ref{eq:2DSPO}) is still demanding since the structure
constants involved are long-ranged as far as the interlayer 
distances are concerned.
In order to render this problem tractable the concept of 'screening'
has been introduced in the middle of nineties \cite{ZDU+95}.
Without going into details, a canonical transformation of the
{\em t}-matrices and the structure constants, which leaves the
Green function invariant, in terms of repulsive scattering potentials
makes it possible to reduce the spatial range of the effective
structure constants. As this
transformation is independent of the real scatterers in the system,
it is very useful to perform self-consistent calculations, 
since it has to be performed for the structure constants only once
at the beginning of calculation. In terms of the 'screened' quantities
Eq.~(\ref{eq:2DSPO}) has exactly the same form.   However, since
$\underline{G}^{pq}(z; \vec{k}_\parallel)$ is now well-localized,
it can be truncated for $|p-q|>n$ at a given $n$ ($\simeq 3$
for FCC and BCC principal facets), which 
in turn implies a block-tridiagonal form for the matrix
${\bf G}(z; \vec{k}_\parallel)$. This blocks are related to the
so-called 'principal layers' containing $n$ subsequent atomic layers. 
By splitting our system into left and right perfect semi-infinite
subsystems, in each of them the scatterers are all identical (bulk),
and into a central region, where the potentials as well as the
orientation of the magnetization can vary, the projection of 
the SPO, Eq.~(\ref{eq:2DSPO}), onto the central region can
be calculated {\em exactly}, i.e. taking into account all the
scattering events to the left and right semi-infinite regions \cite{SUW+94}.
A remarkable feature of the method is that, if only
the layer diagonal blocks of the SPO need to be calculated,
it scales linearly with the size of the central region, 
namely $N$ \cite{God91}. 
Thus it opens the way for investigating systems with inhomogeneities extending 
much beyond the atomic scales such as Bloch Walls.

Let us now turn to the task of performing the Brillouin-zone integration

\begin{equation} \label{eq:BZint}
\underline{\tau}^{nm}(z) = \int_{\Omega_{BZ}} d^2k_\parallel \,
e^{-\imath \vec{k}_\parallel (\vec{R}_\parallel-\vec{R}'_\parallel)}
\underline{\tau}^{pq}(z, \vec{k}_\parallel)  \; ,
\end{equation}
where $\vec{R}_n = \vec{C}_p + \vec{R}_\parallel$ and 
      $\vec{R}_m = \vec{C}_q + \vec{R}'_\parallel$,
while $\Omega_{BZ}$ denotes the volume of the 2D BZ. In Ref. \cite{SUW95}
we described a method for reducing the demand of the above
BZ integration using the symmetry operations of the 
underlying lattice. This is obviously useless for the present case of
the 90$^{\rm o}$ Bloch Wall, since the direction of the magnetization 
rotates from layer to layer, say, from the $x$ axis to the $y$ axis,
therefore the Bloch Wall itself is not invariant under any of the symmetry
operations of the $C_{4v}$ group characteristic to the BCC(001) BZ.

However, it is still useful to note that 
the magnetization direction described by the angle $\phi(z)$ 
in a 90$^{\rm o}$ Bloch Wall perpendicular to the (001) direction 
of a BCC lattice and satisfying the boundary conditions
$\phi(-\infty) = 0$  and $\phi(\infty) = \frac{\pi}{2}$ has
the symmetry property 
\begin{equation}
\phi(z)-\frac{\pi}{4} = -\phi(-z) + \frac{\pi}{4} \; .
\end{equation}
Taking into account also the symmetry of the underlying lattice 
this implies that 
the Bloch Wall is invariant under a $180^{\rm o}$ 
rotation around the axis (110)
which is in three dimensional space represented by the matrix
\begin{equation}
S_{\rm BW}=\left(
\begin{array}{cccc}
0&&1&0\\
1&&0&0\\
0&&0&-1
\end{array}
\right) \quad ,
\end{equation}
whereby the axis of the rotation should cross the $z$ axis at
$z=0$, i.e. for which $\phi(0) = \frac{\pi}{4}$.
Quite clearly, $S_{\rm BW}^{-1}=S_{\rm BW}$. Also evidently,
the 2D square BZ is invariant under $S_{\rm BW}$.

To make use of this symmetry, let $\underline{U}$ be the unitary matrix 
which represents $S_{\rm BW}$ in the
$(\kappa,\mu)$-space and $p'$ denote the layer onto which a
particular layer $p$ is mapped by $S_{\rm BW}$. Then
for the corresponding single-site $t$-matrices we can write

\begin{equation} \label{eq:ttraf1}
\underline{t}^{p'}(z) = \underline{U} \:
\underline{t}^{p}(z) \underline{U}^\dagger
\end{equation}
By introducing
\begin{equation} \label{eq:umat1}
{\bf U}=\{ \underline{U}_{pq} \} \;, \quad
\underline{U}_{pq} \equiv \underline{U} \: \delta_{qp'} 
\end{equation}
or alternatively, by taking the choice of $p'=-p$,
\begin{equation}
{\bf U} =
\left(
\begin{array}{ccccc}
&0&0&\underline{U}&\\
...&0&\underline{U}&0&...\\
&\underline{U}&0&0&
\end{array}
\right)
\end{equation}
it follows that
\begin{equation} \label{eq:ttraf2}
{\bf t}(z) = {\bf U} \: {\bf t}(z) \: {\bf U}^\dagger \; .
\end{equation}

A relationship between the 2D Fourier-transformed structure
constants with arguments $\vec{k}_\parallel$ and 
$\vec{k}'_\parallel = S_{\rm BW} \vec{k}_\parallel$
can also be established using the transformation of the
real-space structure constants

\begin{equation} \label{eq:rsgtraf}
\underline{G}(z; S_{\rm BW} \vec{R}, S_{\rm BW} \vec{R}') =
\underline{U} \: \underline{G}(z; \vec{R}, \vec{R}') \:
 \underline{U}^\dagger \; .
\end{equation}
Therefore, by using Eq.~(\ref{eq:2DFT}) we can proceed as follows

\begin{eqnarray} \label{eq:2Dgtraf1}
\underline{G}^{p'q'}(z; \vec{k}_\parallel) &=& \sum_{\vec{R}_\parallel}
e^{\imath \vec{k}_\parallel \vec{R}_\parallel} \:
\underline{U} \: 
\underline{G} \left( z;\vec{C}_p + S_{\rm BW} (\vec{R}_\parallel
- \vec{R}^p_\parallel + \vec{R}^q_\parallel), \vec{C}_q \right) \: 
\underline{U}^\dagger =  \nonumber \\
&=& \sum_{\vec{R}_\parallel}
e^{\imath \vec{k}'_\parallel \vec{R}_\parallel} \:
\underline{U} \: 
\underline{G} \left( z;\vec{C}_p + \vec{R}_\parallel - 
\vec{R}^{p'}_\parallel - \vec{R}^{q'}_\parallel, \vec{C}_q \right) \: 
\underline{U}^\dagger =  \nonumber \\
&=& \sum_{\vec{R}_\parallel}
e^{\imath \vec{k}'_\parallel 
( \vec{R}_\parallel + \vec{R}^{p'}_\parallel - \vec{R}^{q'}_\parallel )} \:
\underline{U} \: 
\underline{G} \left( z;\vec{C}_p + \vec{R}_\parallel, \vec{C}_q \right) \: 
\underline{U}^\dagger = \nonumber \\
&=& e^{\imath \vec{k}_\parallel \vec{R}^{p}_\parallel} \underline{U} \:
\underline{G}^{pq}(z; \vec{k}'_\parallel) \; \underline{U}^\dagger
e^{-\imath \vec{k}_\parallel \vec{R}^{q}_\parallel} \; ,
\end{eqnarray}
where we defined $\vec{R}^{p}_\parallel = S_{\rm BW} \vec{C}_p -
\vec{C}_{p'}$ and $\vec{R}^{p'}_\parallel = 
S_{\rm BW} \vec{R}^{p}_\parallel$, both being 2D lattice vectors. 
Thus, similar to Eq.~(\ref{eq:umat1}), introducing

\begin{equation} \label{eq:umat2}
\widetilde{\bf U}(\vec{k}_\parallel) =
\{ \widetilde{\underline{U}}_{pq}(\vec{k}_\parallel) \} \;, \quad
\widetilde{\underline{U}}_{pq}(\vec{k}_\parallel) \equiv 
e^{\imath \vec{k}_\parallel \vec{R}^{p}_\parallel}
\underline{U} \: \delta_{qp'} \; ,
\end{equation}
Eq.~(\ref{eq:2Dgtraf1}) can be written compactly

\begin{equation} \label{eq:2Dgtraf2}
{\bf G}(z, \vec{k}_\parallel) =
\widetilde{\bf U}(\vec{k}_\parallel) \:
{\bf G}(z, \vec{k}'_\parallel) \:
\widetilde{\bf U}(\vec{k}_\parallel)^\dagger \; .
\end{equation}
Clearly, the matrix $\widetilde{\bf U}(\vec{k}_\parallel)$ can also
be used instead of ${\bf U}$ in Eq.~(\ref{eq:ttraf2}),
which immediately implies the following transformation
property for the SPO
\begin{equation} \label{eq:2Dtautraf}
\mbox{\boldmath $\tau$}(z, \vec{k}'_\parallel) =
\widetilde{\bf U}(\vec{k}_\parallel)^\dagger \:
\mbox{\boldmath $\tau$}(z, \vec{k}_\parallel) \:
\widetilde{\bf U}(\vec{k}_\parallel) \; .
\end{equation}

As the above relationship makes possible to calculate the SPO only
in half of the BZ when performing the integration in Eq.~(\ref{eq:BZint}), the
computational time and memory storage request of the computer
code can be reduced by a factor of two. Alternatively, in the actual
calculations we have made use of the above symmetry of the Bloch Wall 
by halving the computational demand of the inversion in Eq.~(\ref{eq:2DSPO}),
while keeping all the $\vec{k}_\parallel$ points in the BZ. 
Although, this latter
procedure is almost equivalent to what has been described above,
as it facilitates the tridiagonal shape of the corresponding matrix
\cite{God91}, its use is rather limited to localized schemes
whereas the former one can be regarded to be quite general for
calculations in Bloch Wall problems.

For our present purposes the one-electron Green function, 
Eq.~(\ref{eq:greenf}), can be used to calculate
several quantities of interest such as
the electronic density of states (DOS)

\begin{equation} \label{eq:DOS}
n(z) = - \frac{1}{\pi} Im \int d^3r \; Tr \left( 
{\cal G}(z; \vec{r},\vec{r}) 
\right) \quad (z = \varepsilon + \imath \delta) \; ,
\end{equation} 
where the energy $\varepsilon$ is real, 
$\delta$ is a small imaginary part serving
as parameter of Lorentzian broadening, and $Tr$ denotes trace
of a matrix in the four-dimensional Dirac-space,
the charge density

\begin{equation} \label{eq:rho}
\rho(\vec{r}) = - \frac{1}{\pi} Im \int_{\cal C} d z \: Tr
\left( {\cal G}(z; \vec{r},\vec{r}) \right)  \; ,
\end{equation} 
where ${\cal C}$ is a semi-circle contour in the upper complex semi-plane
starting at the bottom of the valence band, $\varepsilon_{\rm B}$, and ending
at the Fermi energy, $\varepsilon_{\rm F}$,
and the spin-density

\begin{equation} \label{eq:rhospin}
\vec{m}(\vec{r}) = - \frac{1}{\pi} Im \int_{\cal C} d z \: Tr
\left( \beta \: \vec{\Sigma} \: 
{\cal G}(z; \vec{r},\vec{r}) \right)  \; .
\end{equation} 
Furthermore, the charge, the spin-moment and the band-energy 
can be obtained straightforwardly as

\begin{equation} \label{eq:charge}
Q = \int_{\cal C} dz \: n(z) = \int d^3r \: \rho(\vec{r}) \; ,
\end{equation} 
\begin{equation} \label{eq:spin-moment}
\vec{m} = \int d^3r \: \vec{m}(\vec{r}) \; ,
\end{equation}
and
\begin{equation} \label{eq:bande}  
E_{\rm b} = \int_{\varepsilon_{\rm B}}^{\varepsilon_{\rm F}} 
d \varepsilon \: \varepsilon \, n(\varepsilon) =
- \frac{1}{\pi} Im \int_{\cal C} dz \: z \int d^3r \: 
Tr \left( {\cal G}(z; \vec{r},\vec{r}) \right)  \; , 
\end{equation} 
respectively.
From Eq.~(\ref{eq:greenf}) it is quite obvious that the above 
quantities can readily be resolved into components with respect 
to cells (layers)
as we shall show them when we present are results in the
forthcoming sections.

By solving the Poisson-equation for the electrostatic potential
and employing Spin-Density Functional Theory in the Local Approximation 
\cite{VWN80} for the exchange-correlation potential and
exchange field, self-consistent calculations can be performed.
First we carried out self-consistent calculations for BCC bulk iron,
where we have used a theoretical lattice constant of $a$=5.204 a.u. derived by
recent careful full-potential calculations \cite{KSP+00}.
In that calculation we kept the orientation of magnetization along
the (100) easy axis and used 91 $k$-points in the irreducible
wedge of the 2D BZ (IBZ) for the BZ-integrations, Eq.~(\ref{eq:BZint}).
For the energy integrals, Eqs.~(\ref{eq:charge}) and 
(\ref{eq:spin-moment}) we used 16 points along the semi-circle sampled
according a Gaussian-quadrature, while the 
corresponding summations in angular momentum space were subject
to a cut-off of $\ell_{\rm max}$=2. We converged the Fermi energy
to get the corresponding charge (Z=27) to an accuracy of 10$^{-8}$
electrons. Note, that in our present approach, because of the
equilibrium with the left (and right) semi-infinite regions,
this Fermi level has to be also used in the calculations of the (finite-size)  
Bloch Wall. 

As mentioned earlier, we have made use of different approximations 
to calculate the free-energy. 
The first, computationally less demanding, one is based
on the frozen-potential approximation frequently used also
in magneto-crystalline anisotropy calculations (see Ref. \cite{SUW95}
and refs. therein). Briefly,
we used the self-consistent bulk potential and effective field 
magnitude in each layer of the Bloch Wall and 
we set the orientation of the 
exchange field by rotating successively the bulk $t$-matrix 
from layer to layer, Eq.~(\ref{eq:trot}),
according to the prescribed function $\phi_0(z)$. In this case
only one iteration, to calculate the single-particle (band) energy
and the charges, were carried out. As the charge neutrality (number
of particles) is not preserved for the fixed volume of the central
region, the grand-canonical potential has to be considered.
For any thickness $N$ of the Bloch Wall, taking always the difference 
with respect to the ferromagnetic state, this is approximated by
\begin{equation} \label{eq:OmegaN}
\Delta \Omega(N) = \Delta E_{\rm b} - \epsilon_{\rm F} \: \Delta Q =
\sum_p \left( \Delta E_{{\rm b},p} - \epsilon_{\rm F} \: 
\Delta Q_p \right) \;.
\end{equation}
The other approach is a fully self-consistent one. Unfortunately,
the computer power available to us was sufficient for only a 
few such calculations. Therefore, only for the case of $N=60$ 
shall we present and discuss the changes in the electronic
structure and magnetic moments due to the Bloch Wall.

\section{Results}
\subsection{The frozen potential calculations of Bloch Wall energies}
We have calculated the Bloch Wall formation energy 
$\Delta \Omega(N)$ for various values of $N$ 
using the frozen potential approximation. 
Note, that the summation in Eq.~(\ref{eq:OmegaN})
has to be, in principle, taken over all the layers in the system. 
Our calculations show that layers more than about ten layers away from the 
wall do not contribute significantly to the sum although $\Delta E_{p}$ 
and $\Delta Q_{p}$ by themselves differ considerably from zero even far away 
from the wall.
 
As well-known, in the phenomenological theory \cite{Kos86}, 
the exchange energy and the magneto-crystalline energy contribute
equally to the Bloch Wall energy. Experimentally, 
for BCC Fe the magneto-crystalline
anisotropy constant, $K_4$, is found to be 0.3 $\mu$Ryd per
atom\cite{TSE82}, while careful (non-orbital-polarized) 
LDA calculations predict 0.1-0.2 $\mu$Ryd (see Ref. \cite{RSP97} 
and refs. therein and note that for a cubic system 
$K_4=3[E(111)-E(100)]$.)
This, on one hand, implies, that the Bloch Wall 
energy normalized to one layer is expected to be of the same order 
of magnitude as the anisotropy energy and, on the other, we have to calculate 
the Bloch Wall energy to the same 
accuracy as is necessary in magneto-crystalline anisotropy calculations. 

The main difficulty arises from the fact that,
in particular, close to the real energy
axis one has to sample a high number of  
$\vec{k}_\parallel$-points when performing the
BZ-integration in Eq.~(\ref{eq:BZint}). To reduce this problem
we smoothened the energy integrals in Eqs.~(\ref{eq:charge}) and
(\ref{eq:bande}) by the Fermi-function at a finite temperature $T$,
taking into account the poles of it below the contour ${\cal C}$.
Both from an analysis of the integrand and by checking it numerically,
it turns out that the contour can be deformed to infinity in the
upper complex semi-plane. Moreover, only a finite number of 
Matsubara poles, $z_j=\epsilon_{\rm F} + \imath (2j+1) \pi k_{\rm B} T$
$(j=0,1,2,\dots)$, has to be considered. 
Assuming a quadratic $T$ dependence
of $\Delta \Omega(N;T)$ due to the Sommerfeld expansion, 
in order to perform extrapolation to $T$=0
it was necessary to take two different $T$ values only.

In our calculations we choose 300K and 150K for these two
temperatures by using 32 and 40 Matsubara poles with 
1275 and 2926 $\vec{k}_\parallel$-points in the 2D IBZ for $z_0$
(3 and 1.5 mRyd), respectively.
These values were shown to be sufficiently high to yield  converged 
bulk anisotropy energies which we calculated to check the
reliability of our numerical evaluations with respect to other
methods and also to compare to the value that can be deduced 
from the Bloch Wall energies (see Eq.~(\ref{eq:K4}). 
For T=300K and 150K we got $K_4$=0.142 and 0.154 $\mu$Ryd, respectively,
which were extrapolated to $K_4$=0.158 $\mu$Ryd at T=0.
Obviously, this is a very good agreement with the results
of other first-principles calculations.

Let's now turn to the results for the Bloch Wall energies,
associated with an area of size $a^2$, where $a$ is the lattice
constant of our BCC lattice, as a function of thickness $L$,
measured in units of $a$ ($2L/a=N$),
shown in Fig.~\ref{FROZEN} 
for T=150K and T=300K as diamonds and crosses,
respectively. 
The first thing to note is that the calculated points are very well 
fitted by the expression Eq.~(\ref{eq:LINomega}) as drawn
by solid and dashed lines, respectively.
The resulting values for 
$$\hat\alpha \equiv \frac{\alpha \pi^2 a}{4} \quad {\rm and} \quad
\hat\beta \equiv \frac{\beta a^3}{8}$$
are listed in Table~I together what fitted for T=0.

The fitted curves have their minima at 424 and 401 
lattice parameters for T=300K 
and T=150K, respectively. Note that there is a point on the $T=300K$ 
curve which is beyond the minimum and hence we can be said to have 
crossed the scale gap.
The minima are very shallow due to the smallness of the 
anisotropy energy, thus, this is likely to be a feature of 
calculations for all cubic systems.

From the extrapolated values of $\hat\alpha$ and $\hat\beta$
to $T=0$, $K_4$=0.16 $\mu$Ryd and $D$=262 meV\AA$^2$ can be
derived via Eqs.~(\ref{eq:SWstiffness}) and (\ref{eq:K4})). 
By using the expression of $\ell_{\rm BW}$ from the phenomenological
theory we also obtained a thickness of 394 lattice 
parameters for the Bloch Wall.
Several experimental values can be found for $D$, e.g.
$314$ meV\AA$^2$ by Stringfellow \cite{Str68},
$281$ meV\AA$^2$ by Collins et al. \cite{CMN69} or
$280$ meV\AA$^2$ by Mook and Nicklow \cite{MN73}. 
So our value is only slightly below the experimental findings.
The same is true for the comparison
with other theoretical results found from bulk calculations 
\cite{LKAG87,KSK91}. One source of discrepancies between
our and other calculations
is certainly the use of different lattice constants.
It is, however, worth noting that by using the same computer code 
and a spin-flip technique, in the scalar-relativistic limit, a
value of $D$=300 meV\AA$^2$ was calculated \cite{USunpub},
in better agreement with those calculated by others.
This suggest that another source of
the deviation of our present value from those of others 
arises from the relativistic approach we used, that is, the
spin-orbit coupling gives not only rise to the magneto-crystalline
anisotropy, but to some extent influences the spin-spin interaction
parametrized by $D$.

The value of the anisotropy constant is of greater concern. As it is 
generally the case in LDA calculations \cite{TJEW95,RSP97}, 
the experimental value, 0.3 $\mu$Ryd,
is roughly a factor of two bigger than the theoretical results.
However, the important point here is that we find the same value 
for $K_4$ as given above when we determine $K_4$ from 
 bulk calculations demonstrating 
the internal consistency of our calculations.

As implied by Eq.~(\ref{eq:OmegaN})
our calculation allows to resolve the Bloch Wall formation energy
into contributions related to layer. 
This than can be compared to the free-energy density of the 
Ginzburg-Landau theory \cite{Kos86}. Taking the mean value for each
layer and in terms of the parameters introduced above, 
for a linear 90$^{\rm o}$ Bloch Wall this free-energy density reads
\begin{equation} \label{eq:endens}
\epsilon_p= \left\{ \begin{array}{lcl}
\frac{2 \hat\alpha}{N^2} + \hat\beta \sin^2(\phi_p)
&& {\rm if} \quad 1 \le p \le N \\
0 && {\rm anyway}
\end{array} \right. \;.
\end{equation}
In Fig.~\ref{LAYER800} the solid line and crosses in the inset
display the layer-resolved contributions to the Bloch Wall 
energy as calculated by our first-principles method for a
800 layer thick wall and $T$=150K, while
the dashed curve and the diamonds
in the inset are evaluated by using Eq.~(\ref{eq:endens}) with
the the corresponding fitted parameters,
$\hat\alpha$ and $\hat\beta$ (see Table~I.), that is no further
fitting have been used.
Apparently, the coincidence of the two curves is nearly perfect
even in this 'atomic scale' resolution of the Bloch Wall energy.
As is clear from Eq.~(\ref{eq:endens}), in the phenomenological theory
the exchange contribution to the layer resolved energy is 
constant, which equals $2 \hat\alpha/N^2$, inside the wall
and zero outside.  A characteristic deviation from that behavior
is found for the first-principles values near the edge of the wall.  
The exchange contribution to the first layer outside the Bloch wall,
i.e., to layers numbered by 0 and 801 in Fig.~\ref{LAYER800}, is exactly 
half of the above constant value (see inset of Fig.~\ref{LAYER800}).
This is due to the non-local nature of the exchange couplings
not accounted properly within the phenomenological theory.
We note that Bloch Wall formation can be discussed also by using
an effective Heisenberg Hamiltonian with a cubic anisotropy term, 
in which the above feature is readily recovered.        
Small oscillations can also be seen near the edge 
of the Bloch Wall. These are most likely related to Friedel-type  
oscillations, which in homogeneous systems arise due to any imperfections,
as relaxations in the electronic structure.  

\subsection{Selfconsistent calculation}
In order to access reliably 
the electronic structure in the presence of a Bloch wall 
selfconsistent calculations have to be performed.
These calculations are not ground state calculations
because the magnetic moment in every layer is 
forced to point along a prescribed direction.
The proper way to perform such calculations is to introduce a constraining 
field which forces the moment to point along the chosen direction 
\cite{SUW+98}. For our first attempts we ignored the 
constraining field and, instead, took the projection of the magnetic 
moment onto the prescribed direction after every iteration. 

We have performed this calculation for a 60-layer Bloch wall with 
a magnetization profile according to Eq.~(\ref{eq:LINBW}). We allowed 
21 layers on the two sides outside the Bloch wall to relax. 
That is, given the symmetry of the Bloch Wall,  
51 {\it different} potentials were involved in this calculation.
The charging due to the Bloch Wall  per atom for the different layers works out 
to be  smaller than $10^{-7}e$ and, therefore, can be considered
to be zero within the accuracy of our calculations. This means
we observed no charge redistribution (transfer) across the Bloch Wall.
Despite of this fact, as inferred from Fig.~\ref{DOS}, 
the densities of states show characteristic changes for different
layers in the Bloch Wall. Interestingly, the biggest changes are
found at energies with also big peaks in the bulk DOS, which
presumably indicates lifting of some degeneracies, related to the cubic
symmetry, due to the presence of Bloch Wall.

Contrary to the charges, the magnetic moments display a small but
clear deviations from their bulk value.
In Fig.~\ref{MAG} we show the deviation of the moment from the bulk value. 
Obviously, the moments decrease gradually when approaching the
center of the wall. Surprisingly, however, the  
moments outside the wall still quite differ from their bulk value,
approaching it relatively far from the wall only.
Similarly, at the center of the wall one would expect a moment
equal to that calculated for a bulk with (110) hard axis.
This value is shown by dashed line in Fig.~\ref{MAG}.

The component of the moment parallel to the wall but perpendicular 
to the exchange field is shown in Fig.~\ref{MAGY}. 
(One should note that the frame of reference is a local 
one which turns round with the exchange field.)
Not surprisingly, the strong peaks appear at those layers 
which are just outside the Bloch Wall, as these are the layers 
with the most asymmetric neighborhood.
Clearly, these peaks are an artifact due to the prescribed 
magnetization profile which display pronounced kinks at the 
two borders of the Bloch Wall (see Fig. 2). 
The component perpendicular to the wall is smaller than 
$10^{-5}\mu_{\rm B}$ and therefore negligible.

These calculations were performed with 91 
$\vec{k}_\parallel$-points in the two 
dimensional irreducible Brillouin zone.
To achieve convergence roughly 100 iterations have been necessary
(Where one iteration took roughly 14 min on 16 nodes on a T3E).
Clearly, the number of selfconsistent iterations necessary to obtain 
reasonable convergence goes up with the number of layers and we have not 
been able to achieve convergence for thicker 
Bloch walls within the limited CPU time available to us.
Due to the relatively small number of 
$\vec{k}_\parallel$-points used in this calculations 
we do not claim convergence for our results. But we expect the results
to be qualitatively correct. 

\section{Conclusions}
In short, we have presented a first-principles, that is to say 
parameter-free and yet materials specific, description of Bloch Walls
in ferromagnets. As an example we have prescribed the orientation
of the magnetization density, $\vec{M}(\vec{r})$, to evolve
from the easy axis (100) to (010) in BCC iron and calculated its
energy, $E_{\rm BW}$, as a function of the width, $\ell_{\rm BW}$, of
the transition region. To do this we used fully relativistic
spin-polarized Density Functional Theory (DFT) and solved
the Kohn-Sham-Dirac equation by the SPR-SKKR method\cite{SUW95}.
Due to the relativistic description of the electrons, their spin
and orbital degrees of freedom were treated on equal footing and,
hence, the calculation gave a full account of the magneto-crystalline
anisotropy. Consequently, the Bloch Wall energy 
had a minimum. The equilibrium width as well as the full curve,
$E_{\rm BW}(\ell_{\rm BW})$ displayed in Fig. 3. was found to be
in good agreement with available experimental data.

The novel feature of this kind of electronic theory of a Bloch Wall,
as compared to its conventional phenomenological description
\cite{Bloch32}, is that it provides an account of the distortions
of the electronic structure due to the presence of the Bloch Wall
as well as its shape, width and energy. In particular, we have
calculated the variations of the size of the magnetic moment, related to
the local exchange splitting, and the local densities of states
from layer to layer across the transition region. 
These results are displayed in Figs. 5. and 6. Evidently, these
changes are small as expected on the grounds that BCC Fe is
a good moment system. Nevertheless, even in this case they contain
the essential information to describe the scattering
of electrons by Bloch Walls as one needs to do in a study
of magneto-transport \cite{TNN+99,TZM+99,vHSK+98,vHSB99}.

Finally, we note that our calculations scale linearly with
the number of layers, $N$, within the Bloch Wall. Due to
this fact we were able to perform calculations for up to 800 
layers. That is to say, for the first time, we were able
to describe a mesoscopic magnetic defect in fully first-principles
terms. 

\vskip 0.5cm
\centerline{\bf Acknowledgements}

This work resulted from a collaboration partially
funded by the TMR network (Contract No. EMRX-CT96-0089)
and the Hungarian National Science Foundation (Contract No. OTKA T030240 and
T029813). 
One of us (J.S.) would like to thank to
Computational Collaborative Project 9 of UK for financial support.
Most of the computations were performed on the 3TE machine
at the Manchester Computer Center of the EPSRC.

\bigskip
\begin{table}
\begin{center}
\begin{minipage}{7cm}
\begin{tabular}{crc|ccc|ccc}
&T [K] &&& $\hat\alpha$ [$m$Ryd] &&& $\hat\beta$ [$\mu$Ryd] & 
\\ \hline \hline
&300   &&& 64.1                 &&&  0.0356 & \\
&150   &&& 63.2                 &&&  0.0392 & \\ \hline
&0     &&& 62.9                 &&&  0.0404 &  
\end{tabular}
\end{minipage}
\end{center}
\caption{Parameters derived from a least square fit of
the data in Fig.~3 for T=300K and 150K 
to the function Eq.~(\protect{\ref{eq:LINomega}}).
In the last row the corresponding values from a quadratic
interpolation to T=0 are found.}
\end{table}

\begin{figure}
   \begin{center}
   \leavevmode
   \epsfxsize=0.4\textwidth
      \epsffile{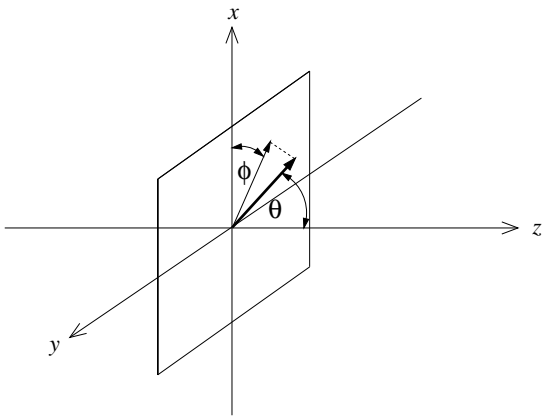}
   \end{center}
\caption{\label{FIG1}
Sketch of the spherical coordinates, $\theta$ and $\phi$,
 as used throughout in the paper.
}
\end{figure}

\begin{figure}
   \begin{center}
   \leavevmode
   \epsfxsize=0.5\textwidth
      \epsffile{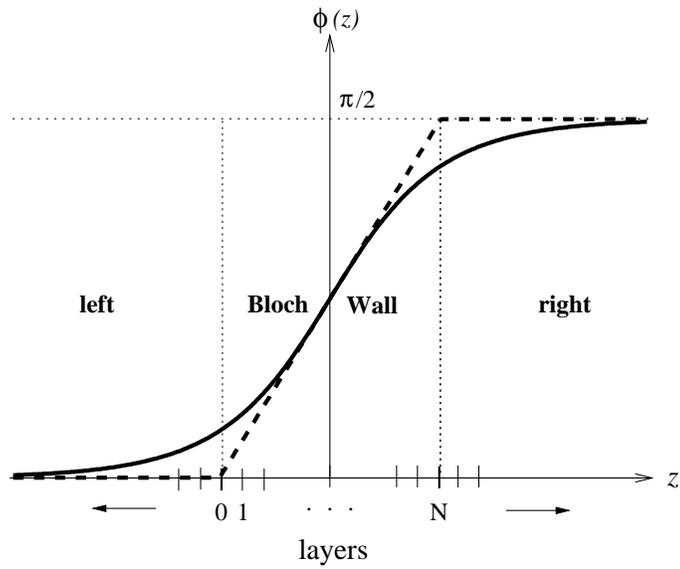}
   \end{center}
\caption{\label{FIG2}
Comparison of the soliton solution, Eq.~(\protect{\ref{eq:soliton}}),
(solid line) and
the linear magnetization profile, Eq.~(\protect{\ref{eq:LINBW}}),
(dashed line) used in our calculations.
The numeration of the layers and the partitioning of the system
into different regions as used in the actual calculations is also
depicted on the picture.
}
\end{figure}

\vskip 2cm

\begin{figure}
   \begin{center}
   \leavevmode
   \epsfxsize=0.5\textwidth
      \epsffile{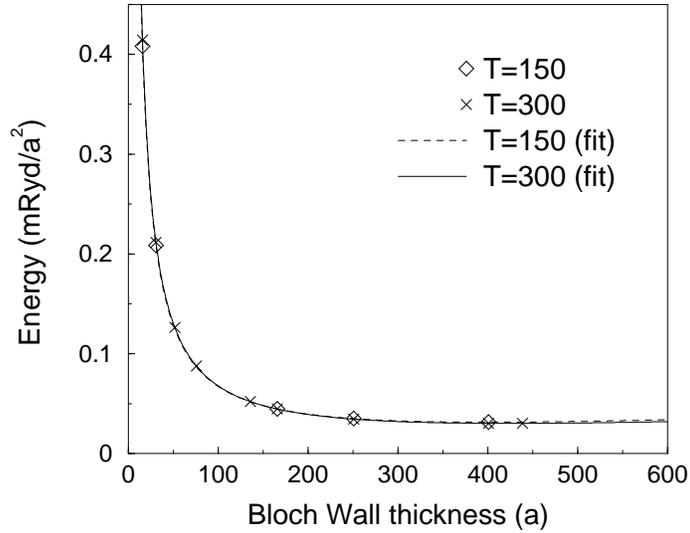}
   \end{center}
   \vspace*{1cm}
\caption{\label{FROZEN}
Bloch wall energy as a function of the Bloch wall
thickness. The diamonds and crosses stand for the results from the
frozen potential calculations for T=150K and T=300K, respectively,
with corresponding fits to the function Eq.~(\ref{eq:LINomega}) 
displayed in order by the dashed and solid lines.
}
\end{figure}

\begin{figure}
   \begin{center}
   \leavevmode
   \epsfxsize=0.5\textwidth
      \epsffile{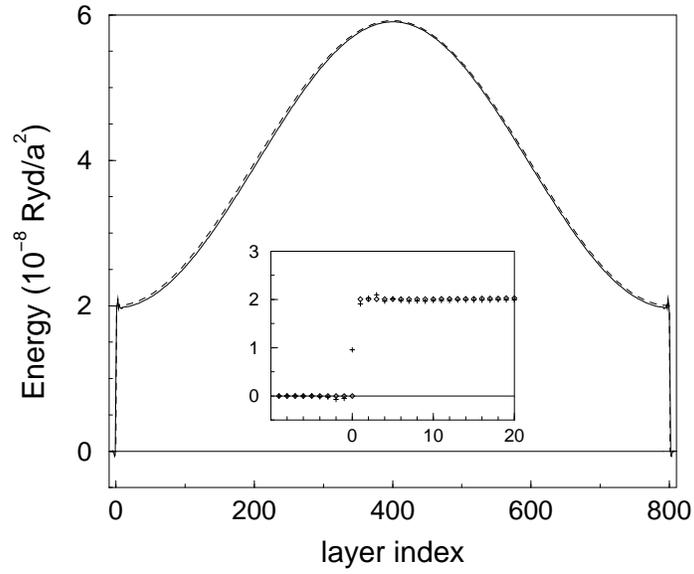}
   \end{center}
   \vspace*{1cm}
\caption{\label{LAYER800}
Layer resolved energies for a 800 layer Bloch Wall.
The full line (crosses in the inset) show the result from the frozen
potential calculation at T=150K. The inset shows
a magnification of the border area.
The dashed line (diamonds in the
inset) show the layer resolved energy contributions according to
Eq.~(\ref{eq:endens}) of the phenomenological theory with
parameters taken from Table~I. 
}
\end{figure}
\vskip 2cm

\begin{figure}
   \begin{center}
   \leavevmode
   \epsfxsize=0.5\textwidth
      \epsffile{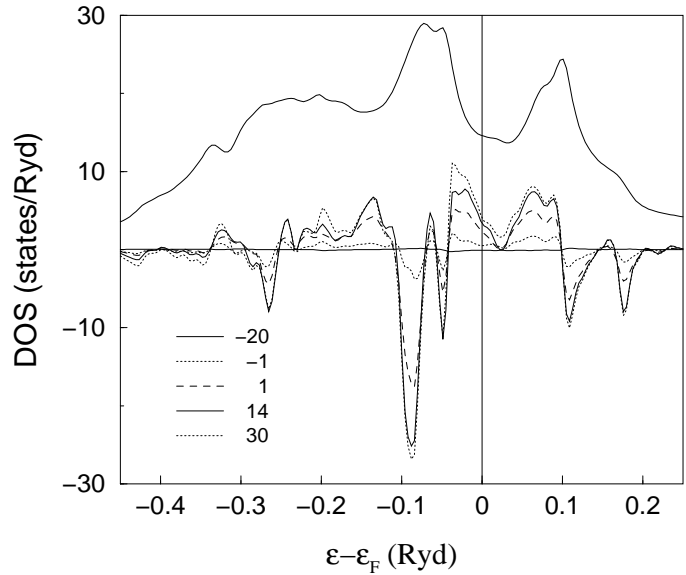}
   \end{center}
   \vspace*{1cm}
\caption{\label{DOS}
The density of states (DOS) for bulk Fe BCC (thick line)
together with their difference from the bulk for several layers in a
60 layer Bloch Wall magnified by
1000. In the legend, negative/positive numbers label layers 
outside/inside the Bloch Wall.
}
\end{figure}

\begin{figure}
   \begin{center}
   \leavevmode
   \epsfxsize=0.5\textwidth
      \epsffile{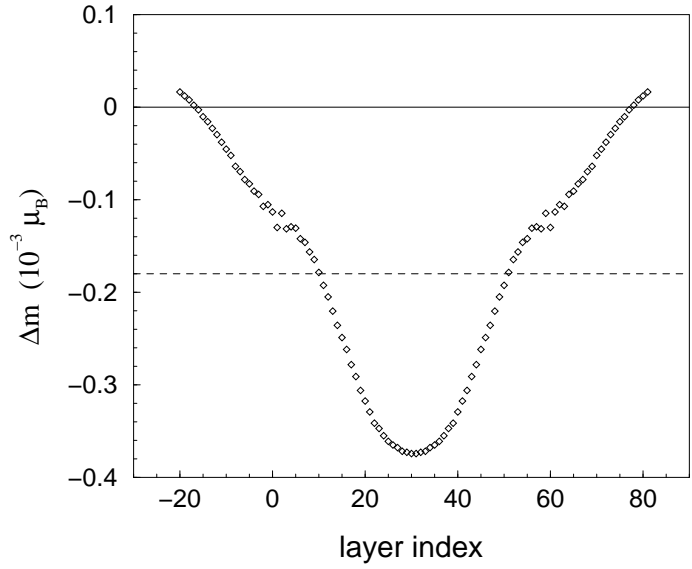}
   \end{center}
   \vspace{0.5cm}
\caption{\label{MAG}
The deviation of the magnetic moment from its
value in the bulk, where it points in the easy direction (100),   
for each layer in a 60 layer Bloch Wall.
The dashed line indicates the moment obtained in a bulk
calculation with the moment along the (110) direction.
}
\end{figure}

\vskip 2cm

\begin{figure}
   \begin{center}
   \leavevmode
   \epsfxsize=0.5\textwidth
      \epsffile{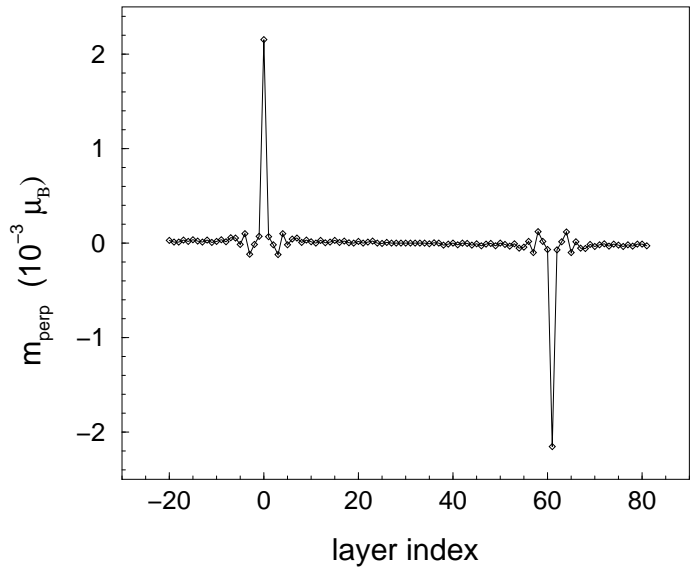}
   \end{center}
\caption{\label{MAGY}
The perpendicular component of the magnetic moment
for each layer in a 60 layer Bloch Wall.
}
\end{figure}


\end{document}